\documentstyle[graphicx,psfig,draft]{mn}

\title[Modification of Dust Grain Structure by Sputtering]
      {Modification of Dust Grain Structure by Sputtering}

\author[M.\,D.\ Gray \& M.\,G.\ Edmunds]
       {M.\,D.\ Gray$^{1}$ \& M.\,G.\ Edmunds$^{2}$\\ 
        $^{1}$ Department of Physics, UMIST,
	       PO Box 88, Manchester, M60 1QD, UK\\ 
        $^{2}$ Department of Physics \& Astronomy,
               Cardiff University, PO Box 913,
               Cardiff, CF24 3YB, UK\\}

\date{Accepted ... .
      Received ... ;
      in original form ...}

\pagerange{000--000}

\begin{document}

\maketitle

\begin{abstract}
We have applied the {\sc SRIM} computer code to study the sputtering of some
likely astrophysical grain materials, and we have shown that selective
embedding of metallic projectiles offers a partial explanation of gas-phase
depletions. We show that supernova shockwaves sweep a significantly larger
mass of interstellar gas per unit time than the shockwaves generated by
outflows in star-forming regions. We apply our sputtering model to the
bombardment levels expected in a supernova shock, and show that net embedding
may dominate over net sputtering, leading to grain growth under some
circumstances, particularly when the bombarding gas is enriched with metals
from the supernova progenitor star. A combination of short cooling times and
net embedding mean that it is possible for a type II supernova to generate
more dust that it destroys, and we conclude that, in general, the sputtering
process often leads to a compositional change in the grain material rather
than simply to grain erosion.
\end{abstract}

\section{Introduction}

The life-cycle of astrophysical dust is not well understood. Whilst certain
constraints can be set on the class of stars which are sources of dust
(Whittet \shortcite{whit92}, Jones \shortcite{jones97})
and additional information can be gleaned from isotopic
compositions of grains \cite{zin98}
we know surprisingly little about the fate
of an average grain from the point where it passes from the stellar wind
of its host star, or supernova, into the wider interstellar medium (ISM).

Of particular interest for models of the life-cycle of dust \cite{mike01} are a set of competing processes which control the
size and mass of the grain, some tending to increase the average grain mass
and others degrading or even destroying the grain. The effect of many of
these processes on grains were considered by Barlow \& Silk \shortcite{bas77}
and Barlow (1978a,b).  
One way to divide up these
processes is to classify them as single particle processes, where a single
gas atom or molecule collides with the grain, and grain-grain processes where
there is a collision between two lumps of solid material. In both cases the
result of the collision, in terms of accretion or erosion of the grain,
depends on the details of the collision, such as the energy, angle, and
composition
of the collision partners. Such processes have been included in sophisticated
models of chemical evolution, which include the development of Galactic
dust properties, for example Dwek \shortcite{dwek98} and references therein.

Single particle collisions with a grain may result in accretion of gaseous
material onto the grain, or in sputtering when the incoming particle has
sufficient energy to eject a grain particle into the gas. In cold molecular
clouds, we can deduce that the single particle processes lead to accretion
of gas molecules onto the grains, leading to an increase in the average
grain size, and mass \cite{dem98}. The key
observations in this respect are signatures
of ices on dust in the spectra of dark clouds \cite{whit96}. Moreover, we can
see that as the
molecular cloud regions become colder, more and more volatile ices are
deposited on the grains, with signatures of non-polar ices, such as
non-polar CO,
appearing instead of the water-ice and other polar ices
common in higher temperature
regions \cite{gibb00}.

Far less is known about the consequences of single particle collisions in the
warmer and more diffuse ISM which surrounds dark clouds. Observations are
sufficient to show that in these regions, grains have lost their ice mantles,
exposing bare core material. There is evidence \cite{whit92} that this is often
of silicate composition, but may also be carbonaceous. It is also quite
possible that refractory organic material acts as a kind of inner mantle
\cite{gal96} protecting a true inner core of silicate. The absence of ice
mantles is often taken to show that grain erosion by sputtering must be an
efficient process, particularly in the warm ISM. Although the thermal energy
of gas atoms in this phase is typically well below the surface energies of
likely grain materials, making thermal erosion impossible, thermal 
and non-thermal
sputtering in supernova shocks is usually considered to be adequate to
supply sufficient gas or plasma particles with the necessary collision
energies (McKee \shortcite{mckee89}, McKee et al. \shortcite{mckeex}).
Indeed, supernova shocks are taken by many authors to be the main
sites of grain destruction in the ISM, for
example, Itoh \shortcite{itoh85}, McKee \shortcite{mckee89}. 
The extent of destructive processes is, however, rather controversial, 
since there is evidence from SN1987A  
\cite{koz91}, \cite{are99} and Cas~A \cite{dunne03} that
supernovae can also produce new 
dust - although its subsequent fate is unknown. Theoretical condensation 
models \cite{tod01} and general consideration of grain abundances 
\cite{mike01} would argue for supernovae being a significant source of
interstellar grains, implying that destruction in these environments 
cannot dominate.

Obviously the main single particle collision partners for grains will be
hydrogen and helium, and in supernova shocks these will probably be in the
form of ions. Heavier ions, which carry more momentum, are more
effective at sputtering than their abundances would indicate. However, an
extensive study \cite{df97} indicates that their total effects are probably
not more than those of the hydrogen and helium, at least for typical Galactic
abundances of metals and an energy spectrum of projectiles which extends well
above the sputtering threshold
(see Section~3.1). More recent work \cite{may00} has studied
the effects of sputtering of typical silicate core materials in considerable
detail, though restricting the bombarding particles to helium and heavier
species.
 In this work, we present a detailed account of the sputtering process, via the Monte-Carlo ion impact
computer code {\sc srim} \cite{zbl85}, as in \cite{may00}, but consider
phenomena not covered by that work: timescales for the release of grain
material, sputtering by hydrogen, sputtering of ice-mantles, and the study
of sputtering as a process of grain modification, rather then the
simple errosive process which is normally assumed.

\section{The SRIM package and parameters}

The {\sc srim} computer package was originally written to solve problems in
nuclear physics, where the penetration of a target by ionizing particles is
to be investigated. The program takes a Monte-Carlo approach, in which the
tracks of individual particles through the target are followed, and the
results from many such independent tracks are accumulated to derive
useful statistics for the target/projectile combination. It should be
emphasized that the individual tracks are completely independent, and assume
a virgin target: that is projectiles which are stopped within the target
are assumed not to change its composition. Part of this work (see Section~3.3
and 3.4)
lifts this assumption by following a set of calculations, each of which
has a target composition modified according to the results of the previous
computation.

The basic physics contained in the {\sc srim} package relates to various
nuclear stopping processes, by which the input projectile loses its original
energy to particles in the target. Secondary processes, or cascades, are
dealt with fully, that is any target particle which recoils with a significant
fraction of the energy of the projectile is then treated as a further
projectile which then also has to lose its energy to further target atoms,
or escape from the target. There are various parameters involved in running
{\sc srim}: the target composition and thickness, projectile type and
energy, projectile angle, and 
various operational flags which control the
form of the output, and the detail in which the projectile tracks are
reported. There are also three energies related to the lattice of the
target: the surface energy, $E_{s}$, the displacement energy, $E_{D}$,
and the lattice binding energy, $E_{B}$. 
The first of these is the energy needed to eject a surface
atom from the target. The second is the energy required to displace an
atom from its lattice site, measured from the bottom of the potential
well in which the target atom can move, to the top of the potential barrier
that must be overcome to leave the site.
Once free of its lattice site, a target atom may be able to escape the
target as a sputtered atom, but if it does escape to infinity, it must
have gained an energy at least equal to the lattice binding energy, relative
to when it was bound in its lattice site. 

Most of the work reported here involves the bombardment,
by protons and heavier ions, of targets which
approximate reasonably to interstellar dust surfaces. We note that {\sc srim}
has already been used to study the effects of sputtering by heavy ions
in the CNO group \cite{df97}, and for more limited results with hydrogen
and helium \cite{tie94}. The versions of the {\sc srim} code used in this
work were {\sc srim-2000.10}, for the comparison with Tielens et al.
\shortcite{tie94}, and
and {\sc srim-2000.40}, for the work on olivine substrates.

\section{Sputtering Results}

The results of this work fall into the following categories: First, we verify
the {\sc srim} method, by comparing our sputtering yields, as a function of
projectile energy and angle, with the work of other authors, both
for {\sc srim} calculations, Field et al. \shortcite{df97}, May et al. 
\shortcite{may00}, and for other methods \cite{tie94}.  Secondly, we consider the effects of selective sputtering
by introducing small amounts of metallic impurities into the grain material;
we discuss whether or not this could help to explain the observed gas-phase
defficiencies in certain elements. Thirdly, we discuss the overall effect
of the bombardment of grain materials by common ions,
including the modification of the
grain material by projectiles which are stopped inside the grain.

\subsection{Comparison with Previous Work}

In Figure~1, we plot the {\it sputtering yield}, $Y(E)$, {\it defined as the number of
target atoms ejected per projectile, expressed as a percentage }, as a
function of projectile energy, for three different target materials. The
projectiles are taken to strike the target at an angle, $\theta$, of zero
with respect to the surface normal. The projectiles are hydrogen nuclei
in all cases, and the energy range is from 2\,eV to 2\,MeV. The target
materials have been chosen to represent three likely populations of
astrophysical grains: graphite and silicate to represent grain
cores, and water-ice to represent grains which have accumulated cold mantles.
In Figure~2, for a more restricted range of projectile energy, we show the
angle-averaged sputtering yield, assuming an isotropic distribution of
projectile trajectories. This average sputtering yield is given by
\begin{equation} \label{eq:1}
\bar{Y}(E) = \int_{0}^{\pi /2} Y(E,\theta ) \sin \theta d\theta
\end{equation}


\begin{figure}
\par
\centerline{
\psfig{file=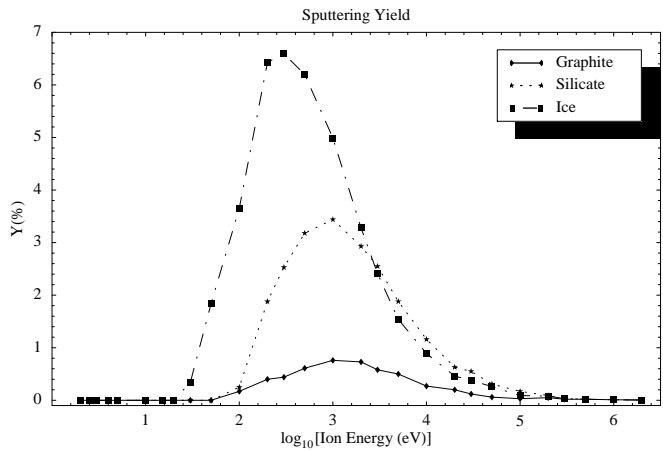,width=5.4cm,angle=270}
}
\par
\caption{
Percentage sputtering yields for hydrogen nuclei, normally incident
upon surfaces of graphite, silicate, and water-ice, as a function of the
projectile kinetic energy. 
}
\label{f:nsputter}
\end{figure}

\begin{figure}
\vspace{0.3cm}
\psfig{file=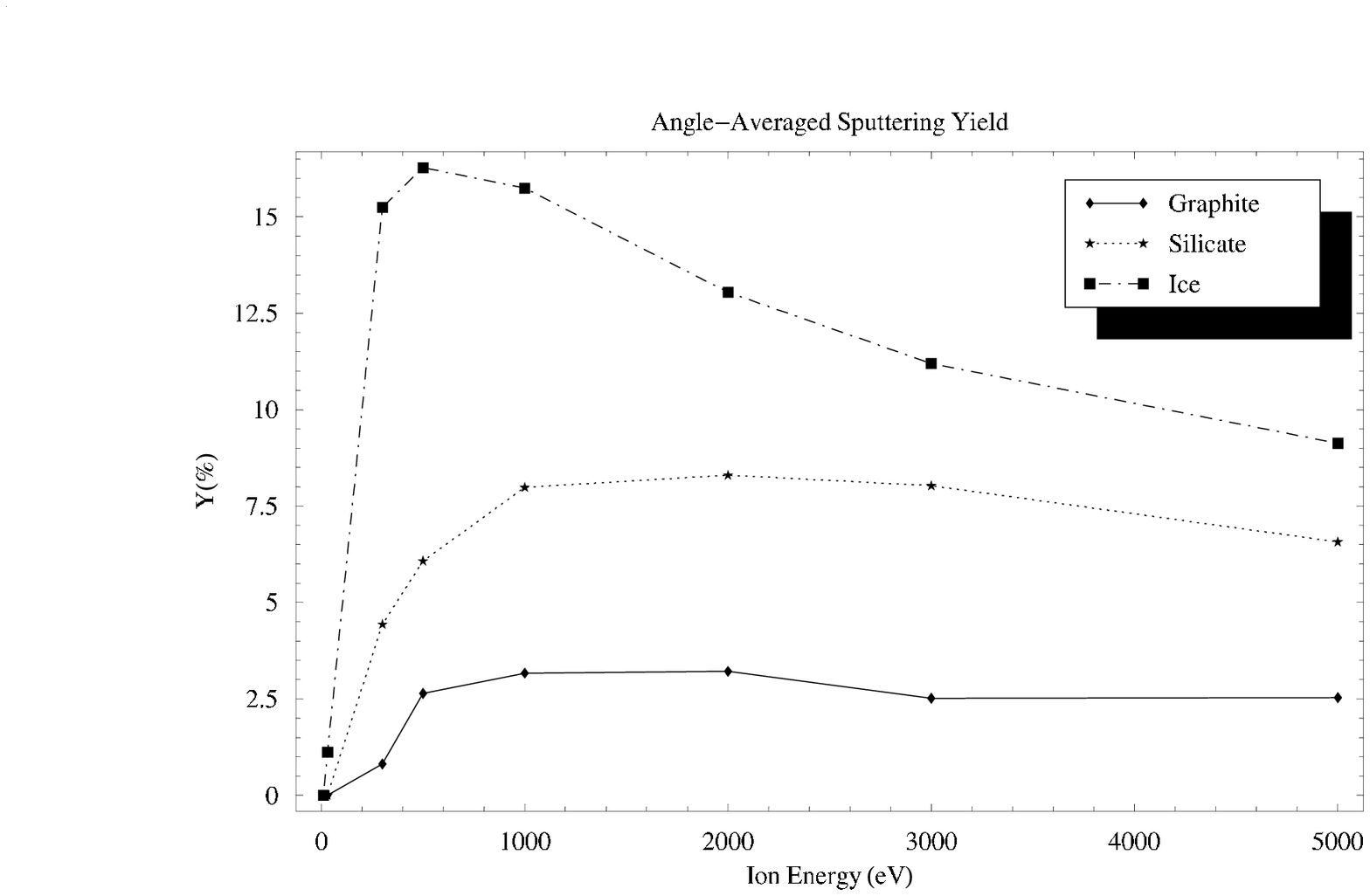,width=5.4cm,bbllx=100pt,bblly=120pt,bburx=382pt,bbury=400pt,angle=270}
\caption{
Angle-averaged
percentage sputtering yields for hydrogen nuclei, incident
upon surfaces of graphite, silicate, and water-ice, as a function of the
projectile kinetic energy. An isotropic distribution of projectile
trajectories has been assumed.
}
\label{f:meansput}
\end{figure} 

The main features of Figure~1 are the following: Sputtering is an essentially
inefficient process; we see that even at the optimum energy for the most
susceptible material (ice) the yield is less then 10\%. We also see that there
is a band of interesting energies. For the target used here, of thickness
100\,nm, we find little sputtering outside the range 30\,eV to 100\,keV.
The lower figure depends only on the surface energy, displacement energy and
lattice binding energy
of the material. On the other hand, the
upper limit depends on the grain size, so that we expect 
significant sputtering damage to persist to higher energies for large grains,
whilst very small grains will become immune to sputtering at lower energies
than in Figure~1. This view of the sputtering process is reinforced by
Figure~2: even with the bias to large angles introduced by the solid-angle
integral in eq.(\ref{eq:1}), no mean sputtering yield exceeds 20\%.

Field et al. \shortcite{df97} and May et al. \shortcite{may00} do not consider
sputtering by hydrogen projectiles, so we compare our results in Figure~1
and Figure~2 to the work of Tielens et al. \shortcite{tie94}, which also
contains some experimental data. For carbon (graphite) the top left-hand
graph of Figure~10 of Tielens et al. shows hydrogen sputtering from a carbon
(graphite)
surface. The peak yield is about 1\% at 350\,eV, and the threshold for
sputtering is around 40\,eV. In our work (Figure~1) we find a peak yield of
about 1\% for impact energies near 1\,keV, and a sputtering threshold energy
near 50\,eV. The hydrogen sputtering in Figure~1 of Jurac, Johnson \& Donn
\shortcite{jur98} has a similar peak yield, but is shifted to slightly
lower energy than the equivalent graph in Tielens et al.
The silicate graph in Figure~10 of Tielens et al. does not
have a plot for hydrogen, but comparing it with the graph for SiC, and assuming
the same order of magnitude difference between H and He sputtering, our figure
of about 3\% for the peak sputtering yield from hydrogen impacts seems
reasonable. For ice, we find a larger discrepancy between our results and
those of Tielens et al.: we calculate a peak sputtering yield for hydrogen ions,
normally incident on ice (see Figure~1), of 6.5\%, but Tielens et al. find
30\%. It is not clear where this difference arises, but we note that Tielens
et al. have scaled their theoretical curves via the free parameter, $K$,
in order to fit experimental data, of which only a small amount is
available. One likely explanation is that neutralization of a
projectile ion, when striking an ice
surface, can lead to electronic sputtering, a process not included in
{\sc srim} \cite{jur98}. Another is that that there are many allotropes of
ice, each with its own lattice binding energy. However, as we do not consider
ice further in this work, but concentrate on silicate cores, the 
discrepancy noted above is not particularly important here.

For comparison with May et al. \shortcite{may00} we choose an olivine
target, and look at the sputtering yield of silicon when the olivine,
MgFeSiO$_{\rm 4}$, is bombarded by oxygen and iron projectiles (see Figure~2
of May et al). When we reverted our {\sc SRIM} code to include the
restrictions imposed by their earlier version, the results were the same to
within the accuracy with which we could read their Figure~2 ($\sim 5$\%). For
example, at $100$\,eV with oxygen projectiles, we obtained a yield of Si of
$0.018$; the marker on Figure~2a of May et al. at $100$\,eV is clearly just
below $0.02$. For the same graph, at $60$\,eV, we find a yield of
Si of $0.00204$, whilst the graph value at this energy is almost 
exactly $0.002$. This excellent agreement was found in spite of a
small difference in the density used for the
olivine (we used $3.81$\,g\,cm$^{\rm -3}$ \cite{CRC} instead of the
$3.84$\,g\,cm$^{\rm -3}$ used by May et al.), although the yield is
expected to be quite insensitive to the density: Field et al.
\shortcite{df97} state that a $5$\% change in sputtering yield requires
a change in density as large as $20$\%.
Note that we have used the fractional definition of the
yield here, as in May et al., rather than the percentage definition used
elsewhere in this work. The principal restriction of the older version of
{\sc SRIM} used by May et al. is that it allowed only a single value for each
of the energy parameters, $E_{s}$, $E_{D}$ and
$E_{B}$, defined in Section~2. Therefore, they used a weighted average over
all the constituents of each target.
In our work, we used the more recent {\sc srim-2000.40} to study the
olivine target, and this allows separate values
of $E_{s}$, $E_{D}$ and $E_{B}$ to be chosen for all
species in the target. We used the values supplied by 
{\sc srim-2000.40} which appear in Table~1 above. None of the individual
species energies supplied by {\sc SRIM} are as high as the blanket averages
used in May et al. It is therefore not surprising that, in our work on olivine,
we find significant sputtering of Si by
O, Fe, and other projectiles, at $30$\,eV. When using the species-specific
energy parameters, we obtain a fractional yield of silicon, 
at 100\,eV, of $0.055$ for the oxygen
projectile and $0.034$ for the iron projectile. Below this energy, large
descrepancies appear between our data and May et al. \shortcite{may00}. 
Fortunately, we are considering a higher
energy regime than May et al., so the precise energy of the sputtering
threshold is less important.

\begin{table}
\caption{The Energy Parameters $E_{s}$, $E_{d}$ and
$E_{B}$ for Olivine}
\begin{tabular}{@{}lrrr@{}}
\hline
Element   &  $E_{s}$ & $E_{D}$ & $E_{B}$    \\
          &    eV    &    eV   &   eV       \\
\hline
   Mg     &  1.54    &  25.0   &  3.0       \\
   Fe     &  4.34    &  25.0   &  3.0       \\
   Si     &  4.7     &  15.0   &  2.0       \\
   O      &  2.0     &  28.0   &  3.0       \\
\hline
\end{tabular}
\end{table} 

In Figure~3, Figure~4, and Figure~5 we show more detail in the sputtering
yield, including, in these figures, the dependence on the angle of 
incidence of the projectile.
Figure~3 shows the results for a carbonaceous
grain (graphite), Figure~4 shows analogous results for a silicate material
(quartz), and data for water-ice are depicted in Figure~5. All the surfaces
exhibit a qualitatively similar behaviour: higher angles of incidence
increase the sputtering yield for a given energy, until an optimum angle,
usually in the range 75-88 degrees. At higher angles, the number of 
backscattered projectiles becomes so large that sputtering efficiency is
lost. Only near the optimum energies for the most susceptible material
(ice) does the sputtering yield at the optimum angle exceed 30\%. The 
form of the angular dependence in Figure~3 to Figure~5 can be compared to
Figure~2 of Jurac et al. \shortcite{jur98}, noting that the latter results 
are for a helium projectile.

\begin{figure}
\vspace{0.6cm}
\psfig{file=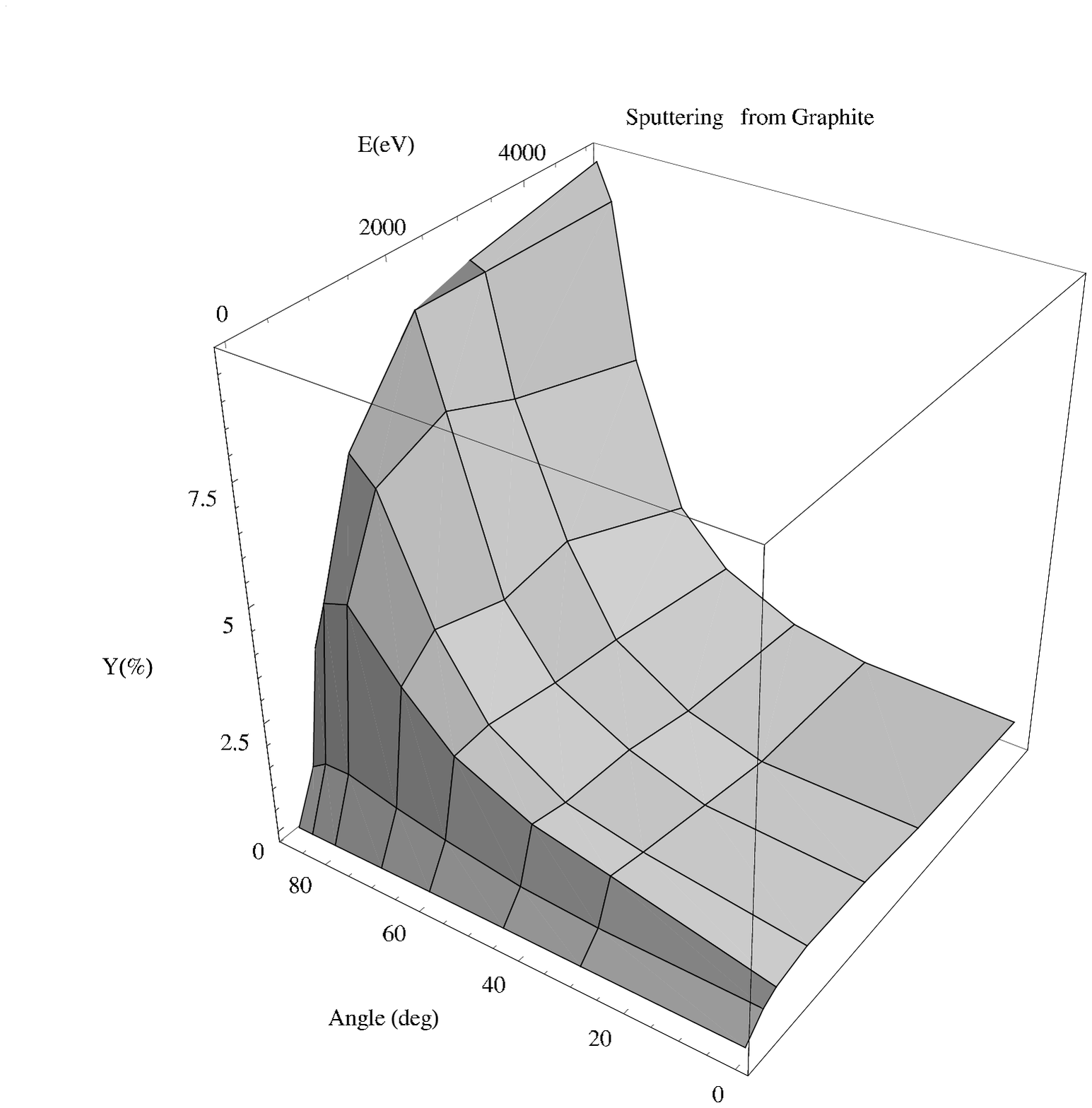,height=10.0cm,bbllx=80pt,bblly=350pt,bburx=462pt,bbury=850pt,angle=0}
\caption{
Percentage sputtering yields for hydrogen nuclei incident
upon graphite, as a function of the
projectile kinetic energy and angle of incidence. An angle of 90 degrees
is parallel to the target surface. 
}
\label{f:gsputter}
\end{figure}

\begin{figure}
\vspace{0.6cm}
\psfig{file=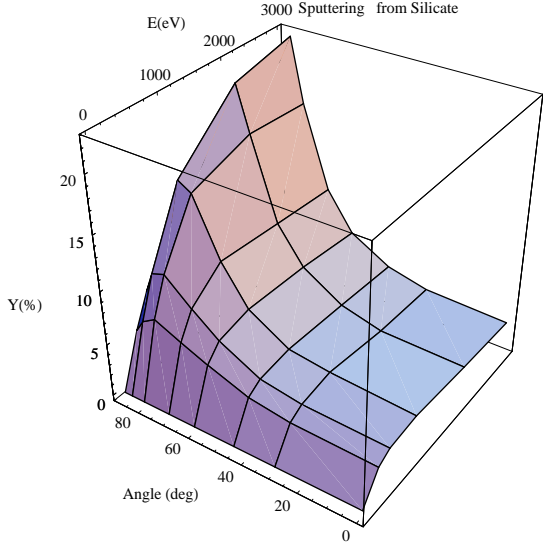,height=10.0cm,angle=0}
\caption{
Percentage sputtering yields for hydrogen nuclei incident
upon crystalline silicate (quartz), as a function of the
projectile kinetic energy and angle of incidence. An angle of 90 degrees
is parallel to the target surface. 
}
\label{f:ssputter}
\end{figure}

\begin{figure}
\vspace{0.6cm}
\psfig{file=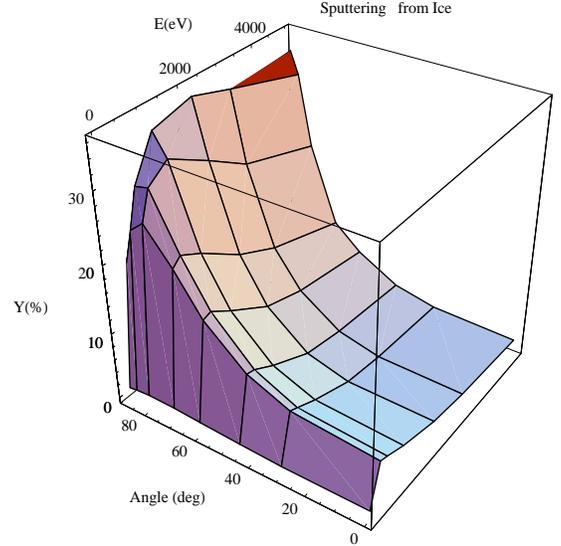,height=10.0cm,angle=0}
\caption{
Percentage sputtering yields for hydrogen nuclei incident
upon water-ice, as a function of the
projectile kinetic energy and angle of incidence. An angle of 90 degrees
is parallel to the target surface. 
}
\label{f:isputter}
\end{figure}

A conclusion that we share with Tielens et al. \shortcite{tie94} is that
thermal sputtering is rarely of importance in the quiescent ISM. Only
the hot, or coronal gas, phase of the ISM has sufficient temperature to
place ions above the sputtering threshold of typical grain materials; this
hot gas is, however, too diffuse to erode grains within a reasonable
time. Sputtering therefore proceeds rapidly only in shocks, in which some
of the particles at least are subject to non-thermal acceleration
mechanisms and/or thermal sputtering at much higher densities than exist
in the coronal gas. The two likely candidates for shock-sputtering of grains
are supernova shocks, for example Tielens \shortcite{tie94},
Jenkins \shortcite{jenk89}, McKee \shortcite{mckee89},
and the C-type shocks fronting molecular outflows
in star-forming regions, as considered by Field et al. \shortcite{df97} and
May et al. \shortcite{may00}. Which of these shock processes is
dominant?

Thermal sputtering is out of the question in cold molecular clouds. Therefore,
Field et al. \shortcite{df97} and May et al. \shortcite{may00} introduce
non-thermal sputtering via a model of a C-type (continuous) MHD shock, of
modest speed ($20$-$45$\,km\,s$^{\rm -1}$), moving through a molecular medium.
The post-shock gas does not
reach temperatures sufficient to ionize most components of the gas, so neutrals
remain abundant, and sputtering results from a large drift velocity between
the neutral and ionized fluids. The grains are taken to be part of the latter
fluid; see the discussion in Weingartner \& Draine \shortcite{wei01} for
the range of possible charge-states that are accessible to astrophysical
dust grains: both positive and negative charging is possible. We now 
calculate the likely sweeping rate of the ISM by shocks of this type, assuming
that they are produced by bipolar outflows, associated with star-formation.
Bontemps et al. \shortcite{bon96} derive an expression for the `momentum
flux' in an outflow, based upon observations. This flux is related to the
bolometric luminosity of the source, which is, in turn, assumed to be produced
by accretion during the outflow phase. Combining expressions for the momentum
flux and accretion power from Bontemps et al. yields
\begin{equation} \label{eq:mflux}
F(M) = 10^{-5.6} \log_{10} \left(
\frac{GM\dot{M}_{acc}}{3R_{\odot}L_{\odot}}
                        \right)^{0.9}
\end{equation}
where the `momentum flux', $F(M)$, is measured in units of solar masses per
year times an outflow line-width in km\,s$^{\rm -1}$, and $\dot{M}_{acc}$ is
the accretion rate. Assuming a timescale for the outflow phase of
$\tau_{0}=\tau_{5} \times 10^{5}$\,yr \cite{sar96}, where $\tau_{5}$ is 
the outflow lifetime in units of $10^{5}$\,yr, and a 
velocity width of $v_{50} \times 50$\,km\,s$^{\rm -1}$, the
rate at which ISM material is shocked by the outflow is given by
$10^{5} \tau_{5} F(M) / v_{50}$, where $v_{50}$ is the velocity
width in units of $50$\,km\,s$^{\rm -1}$, or
\begin{equation} \label{eq:shockrate}
\dot{M}(M) = \frac{10^{-5.6}}{50 v_{50}} \left(
   \frac{GM^{2}}{9.46 \times 10^{12} R_{\odot}\tau_{5}L_{\odot}}
                                     \right)^{0.9}\;\;\;M_{\odot}\,yr^{-1}
\end{equation}
where we have
assumed that the star accretes its own mass during the time $\tau_{0}$, so
that $\dot{M}_{acc}=M/\tau_{0}$. We note that $v_{0}$ is essentially twice
the shock velocity of the outflow. Assuming a modified
power-law initial mass function (IMF), with an index of $1.35$ for
stars below $0.5$\,M$_{\odot}$, and the classical Salpeter index of
$2.35$ for stars of 
greater mass \cite{kit87}, we find that, for stellar mass limits of
$0.08$\,M$_{\odot}$, and $100$\,M$_{\odot}$, the mean stellar mass
is $0.5253$\,M$_{\odot}$ and the mean sweeping rate, found from
\begin{eqnarray} \label{eq:sweep}
\bar{\dot{M}}  & = & \frac{2\times 10^{-7.6}}{v_{50}}
                \left(
\frac{G}{9.46\times 10^{12}\tau_{5}R_{\odot}L_{\odot}}
                \right)^{0.9} \nonumber \\
& \times &
                \left[
0.2367M_{\odot}^{0.35} \int_{0.08M_{\odot}}^{0.5M_{\odot}} M^{0.45} dM
                \right. \nonumber \\ 
& + & \left.
0.1184M_{\odot}^{1.35} \int_{0.5M_{\odot}}^{100M_{\odot}} M^{-0.55} dM
                \right]\;\;\;M_{\odot}\,yr^{-1}
\end{eqnarray}
is $5.78 \times 10^{-6}/(v_{50}\tau_{5}^{0.9})$\,M$_{\odot}$\,yr$^{\rm -1}$. We
note that if the
IMF were extended to brown-dwarf masses, the mean stellar mass, and mean
sweep-rate would be lower than calculated here. Finally, the amount of
ISM swept by the outflow of an average star, during its formation, is
found by multiplying the sweeping rate by $\tau_{0}$, giving
\begin{equation} \label{eq:flowmass}
M_{out} = 0.578 \tau_{5}^{0.1} / v_{50}\;\;\;M_{\odot}
\end{equation}
where we note that the dependence on the outflow lifetime is very weak.

We compare the swept mass of ISM from eq.(\ref{eq:flowmass}) to a
figure for supernovae,
derived by McKee \shortcite{mckee89}, of $6800$\,M$_{\odot}$ before the
shock decelerates to $100$\,km\,s$^{\rm -1}$. Using the same IMF as used
above for the outflows, we find that the fraction of stars formed as
supernova progenitors ($M>8$M$_{\odot}$) is $0.0051$. Assuming that the
modern supernova rate is equal to the formation rate of such stars, then
the relative efficiency of shocking interstellar gas by supernovae and
outflows is in the 
ratio $0.0051 \times 6800 \times v_{50} / (0.578 \tau_{5}^{0.1})$, which 
is equal to $600v_{50}/\tau_{5}^{0.1}$.
Supernovae therefore shock about $600$ times more material than star-forming
outflows, so whilst the outflows can be locally very important, most 
sputtering occurs in supernova shocks unless the sputtering efficiency
in outflows is vastly more efficient than in supernovae; this is unlikely
to be the case.

As a check on the above calculation, we compare observational estimates of
the current supernova and star-formation rates, which are independent of
any assumed IMF. McKee estimates the effective supernova rate (for supernovae
which interact strongly with the ISM, which
effectively excludes type~Ia supernovae, and allows for
correlations due to multiple supernovae in clusters) to be about $1$ event per
$120$\,yr. We compare this with estimates of the Galactic star-formation rate
by Boissier \& Prantzos \shortcite{bois99}, who estimate a rate of
$3 \times 10^{-9}$\,M$_{\odot}$\,yr$^{\rm -1}$\,pc$^{\rm -2}$ for
the solar neighbourhood.
From Figure~2 of Boissier \& Prantzos, we can see that the star-formation
rate does not reach ten times this value at any radius in the Galactic disc.
Adopting a Galactic radius of $15$\,kpc, and using the mean stellar mass,
calculated earlier, of $0.5253$\,M$_{\odot}$, we find a star-formation
rate of $4.0$\,yr$^{\rm -1}$. The ratio of the effective supernova rate
to the total star-formation rate is therefore $0.002$, compared with the
$0.0051$ calculated on the basis of our IMF. These figures are consistent
if we adopt the lower figure, because the value of $0.0051$ is not 
corrected for the effectiveness of the supernovae. With the lower figure,
the supernovae are still about $240$ times as effective at
shocking interstellar gas then star-forming
outflows.

In the light of the analysis above,  
the model we adopt in the present work is very different from
May et al. \shortcite{may00}. We consider a J-type supernova blast-wave moving
at over $100$\,km\,s$^{\rm -1}$, which provides both 
thermal and non-thermal sputtering
projectiles. The post-shock gas is ionized by the supernova shock, so
the projectiles and grains are both part of the charged fluid, and the
ion-neutral
drift velocity mechanism is not applicable. In the present work, we
ignore betatron acceleration of the grains and projectiles, and
also the fact that a grain will initially have a large initial velocity
relative to the gas until its motion is thermalized. We divide
the post-shock gas into a thermal fraction, which has an energy
distribution function given by the Maxwell-Boltzmann form,
\begin{equation} \label{eq:maxboltz}
p(E) = \frac{2}{\pi^{1/2}}
       \frac{E^{1/2}}{(kT_{2})^{3/2}}
       e^{-E/(kT_{2})}
\end{equation}
rather than a skewed Maxwellian (used, for example, in Barlow \shortcite{barlow78a}),
and a small non-thermal fraction, which enters a first-order Fermi
acceleration process above some injection energy, $E_{inj}$. We use the
results of a diffusive shock model \cite{berez99} to obtain our
non-thermal energy distribution functions
and the post-shock conditions. The crucial difference between these models
and simpler test-particle calculations is that penetration of the 
pre-shock gas by energetic particles leads to energy loss from the shock:
there is no longer a simple limiting compression ratio for the shock, and
it can become very large. However, for accelerated particles which have
not reached relativistic energies, the power-law spectrum of the 
non-thermal distribution is not different to the test-particle case
\cite{berez99}. The power-law energy spectrum of ions of species $j$,
derived from Berezhko \& Ellison, is
\begin{equation} \label{eq:espec}
p(E) = \frac{9\eta \lambda^{3/2} \gamma_{g}^{3/4} k^{3/4} T_{2}^{3/4}}
            {2^{3/4} 4 r_{sub}}
\left( \frac{m_{j}}{\bar{m}} \right)^{3/4} 
             E^{-7/4}
\end{equation}
where $\eta$ is the injection fraction of particles into the non-thermal
mechanism, $\lambda$ is the ratio of the injection momentum
to the thermal momentum $m_{H}c_{s2}$, $\gamma_{g}$ is
the ratio of specific heats in the thermal gas, $\bar{m}$ is the mean
particle mass, and $m_{j}$ is the mass of species $j$. The temperature in
the post-shock gas is $T_{2}$, and the adiabatic sound speed in the
post-shock gas is $c_{s2}$.
The power-law index results from taking the
compression-ratio in the sub-shock, $r_{sub}$, to
be $3$ (see Berezhko \& Ellison
\shortcite{berez99}). A crucial quantity is the post-shock temperature,
$T_{2}$. We adopt the most physically realistic case which allows for
loss of energy from the shock, allowing the overall compression ratio to
rise above its classical limit of $4$, but we allow the compression ratio
to be limited by the dissipation of Alfv\'{e}n waves, so it cannot rise
without limit as the Mach number tends to infinity. The post-shock
temperature is therefore given by
\begin{equation} \label{eq:pstemp}
T_{2}=1.9 \times 10^{5} u_{100}^{5/4} v_{A10}^{3/4} \;\;\;K
\end{equation}
where $u_{100}$ is the shock speed in units of $100$\,km\,s$^{\rm -1}$, and
$v_{A10}$ is the Alfv\'{e}n speed in the unshocked gas in units of
$10$\,km\,$s^{\rm -1}$. The thermal and non-thermal spectra are matched at
an injection energy which is assumed to be species dependent: Berezhko \&
Ellison consider only hydrogen, and specify the injection
to thermal momentum ratio as $\lambda = 4.3$. 
We assume that this momentum ratio
holds for all species, such that the injection energy is
$E_{inj}(j) = p_{inj}^{2}/(2m_{j})$. By demanding 
equality of eq.(\ref{eq:maxboltz}) and
eq(\ref{eq:espec}) at the injection energy, we calculate an injection fraction
of 
\begin{equation}
\eta = 1.85 \times 10^{-5} (\bar{m} / m_{j})^{4/3}
\end{equation}
An important consequence of the mechanisms used in the present
work is that, compared to May et al. \shortcite{may00}, hydrogen and helium
are much more important: in the thermal distribution particles have the
same energy, not the same speed, so impact energy is not weighted to
projectiles of higher mass. Also, in the non-thermal mechanism, the injection
energies are higher, and the injection fractions smaller, for ions of
higher mass, reducing their importance relative to light ions. Hence, in
contrast with
May et al., we cannot ignore
hydrogen sputtering in the present work.

\subsection{Selective Sputtering of Impurities}

Certain metallic elements, notably Ca, Ti, Co \& Ni, are observed to be very
under-abundant in the gas-phase of the ISM
\cite{sav96}. It has long been known that gas-phase metallicity correlates
with gas velocities \cite{bas77}, suggesting that shocked gas loses some
of its grain material back to the gas.
We test here, the idea that such
observations may be explained, at least in part, by a selective propensity
of these elements to resist sputtering, relative to the bulk grain material,
or alternatively, of a strong tendency to be stopped, and captured within
a grain when acting as a projectile. These processes are likely to be
complementary to the lower energy differential surface adsorption proposed
by Barlow \& Silk \shortcite{bas77} and Barlow \shortcite{barlow78b}.

The first possibility, a strong resistance to sputtering, has been tested
by admixing each of the grain materials with small percentages of selected
elements as impurities before carrying out the sputtering calculations. We
note that some degree of non-stoichiometric sputtering is expected, with
the more volatile components of the target likely to experience a greater
degree of sputtering \cite{tie94}. Initial calculations were carried out
with the {\sc srim-2000.10} code, in which the energy parameters. $E_{s}$,
$E_{D}$, and $E_{B}$ are not species-specific. These results were negative:
that is, the fraction of impurity atoms
sputtered from the target was not substantially different from the initial
fractional abundance of the same element in the target. In case this
negative result was due to the uniformity of the energy parameters, we also
studied selective sputtering from the olivine target using the more
versatile {\sc srim-2000.40} which allows separate values of the energy
parameters to be set for the Mg, Si, Fe, and O-components of the olivine. We
note that of these consistuents, Mg, Fe and Si are underabundant in the gas
phase relative to oxygen, with Fe being the most extreme of the three and
Si, the least extreme \cite{sav96}. Our results for this case are tabulated
in Table~2. We show angle-averaged total sputtering yields for each
element in the olivine at four different energies. The figures for oxygen
have been divided by four to allow for its stoichiometric abundance in
olivine. The total sputtering yields have been calculated by summing over the
abundance-weighted contributions of the ten commonest Galactic species
(see Appendix~1).

\begin{table}
\caption{Total Sputtering Yields for Olivine Constituents}
\begin{tabular}{@{}lrrrr@{}}
\hline
Element   & $\bar{Y}_{30eV}$ & $\bar{Y}_{100eV}$ & $\bar{Y}_{300eV}$ &
$\bar{Y}_{1keV}$\\
          &    \%            &     \%             &      \%           &
   \%           \\
\hline
   O      & $4.54\times 10^{-4}$ & $0.375$  & $4.010$ & $3.140$ \\
   Mg     & $6.24\times 10^{-4}$ & $0.440$  & $4.576$ & $3.761$ \\
   Fe     & $3.29\times 10^{-4}$ & $0.224$  & $2.913$ & $1.861$ \\
   Si     & $7.75\times 10^{-4}$ & $0.234$  & $2.810$ & $1.699$ \\
\hline
\end{tabular}
\vspace*{1mm}
\noindent \\
Notes for Table~2.\\
Values for oxygen have been divided by four to allow for its
stoichiometric ratio in the target.
\end{table}

The results in Table~2 do show differences between the amount of sputtering
for each element. Close to the threshold for sputtering, Si has the 
highest sputtering yield, which reflects its low displacement energy (see
Table~1). At higher energies we find an approximate binary divide with
oxygen and magnesium more likely to be sputtered than the iron and silicon.
This cannot explain the observed sequence of gas-phase depletions however: we
expect magnesium to be intermediate between iron and silicon, which is
not the case in Table~2, where Mg is more easily sputtered than oxygen
at all but the lowest energies.

The second possibility, a strong tendency to be captured by a grain, looks
far more promising as an explanation for gas-phase depletions. The basic
argument is that an atom from one of the depleted elements, striking a
grain as a projectile, is likely to be stopped within the grain, and
trapped, at a depth characteristic of the grain material, and the impact
energy (see Section~3.3). In the event that it does cause sputtering, it is likely not to
eject an atom of its own kind, or one of similar rarity, but an atom of the
basic grain material: hydrogen, oxygen, carbon, silicon, or perhaps iron.
We therefore expect a steady loss of such atoms from the gas-phase until
we reach an equilibrium state, where the rate of stopping in grains, for
a particular element, becomes equal to the rate of return by the dominant
grain destruction process. Atoms of the heavy elements are more likely to
be stopped inside the target, with a lower probability of backscattering than
hydrogen and helium. In Table~3, we show the backscattered fraction for
projectiles of various types, when incident on an olivine target at $0.1$ and
$1$\,keV.

\begin{table}
\caption{The backscattered fraction, $f_{B}$, of ions of various projectile
ions normally incident on an olivine target at $0.1$ and $1$\,keV}
\begin{tabular}{@{}lrr@{}}
\hline
Element   &   $f_{B}$ ($100$\,eV) & $f_{B}$ ($1$\,keV) \\
\hline
   H      &    0.3272  &  0.1874            \\
   He     &    0.2387  &  0.1845            \\
   C      &    0.0587  &  0.0582            \\
   N      &    0.0451  &  0.0476            \\
   O      &    0.0346  &  0.0381            \\
 Ca$^{*}$ &    0.0030  &  0.0072            \\
   Ti     &    0.0002  &  0.0031            \\
   Co     &    0.0000  &  0.0008            \\
   Ni     &    0.0000  &  0.0008            \\
   Cr     &    0.0001  &  0.0019            \\
   Fe     &    0.0000  &  0.0011            \\
   Mg     &    0.0166  &  0.0200            \\
   Si     &    0.0107  &  0.0151            \\
\hline
\end{tabular}
\vspace*{1mm}
\noindent \\
Notes for Table~3.\\
$^{*}$ Calcium has the highest gas-phase depletion.
\end{table}

The data in Table~3 show that very few ions of the elements with large
gas-phase depletions get backscattered compared with hydrogen, helium and
the C,N,O group of elements. Ions from the depleted group (from Ca onwards
in Table~3) are very likely to be stopped inside the the target, and
become part of it. The
implication is that this is a viable mechanism for explaining interstellar
depletion, although we concede that Ca would not be the most depleted
element on the the basis of these data.


\subsection{Modification of Grain Material}

The `classical' sputtering yield effectively treats the sputtering process
as negative accretion, with a grain radius, $a$, that declines according to the
equation
\begin{equation}
\frac{da}{dt} = - \frac{Y\bar{m}nv}{4\rho}
\end{equation}
where $\bar{m}$ is the mean mass of grain atoms, $n$ is the number density
of projectile atoms in the gas-phase, $v$ is the relative velocity of
projectile and grain, and $\rho$ is the density of the grain.
It is a consequence of the results presented above, that this view of
sputtering cannot be correct because, given that the projectile atoms do
not immediately escape the material, we find a steady increase in the number
of grain atoms with time. We discuss below whether it is reasonable to
expect stopped projectiles to escape the grain material rapidly compared
with other evolutionary timescales. For bombarding gas with normal Galactic
abundance, most of the deposited particles will be light, but we investigate
an alternative evolution for sputtered grains in
supernova remnants in Section~4, where
atoms of the C,N,O group and heavier species have very high abundances relative
to typical interstellar gas. 


First, we plot the stopping probabilites for the standard 100\,nm grain,
with hydrogen projectiles, for the carbon, silicate and ice materials. The
results are shown in Figure~6. We can see immediately that apart from a
high-energy cut-off, dictated by the size of the grain, the fraction of
projectile atoms stopped within the grain is typically above 0.6. This
fraction becomes smaller at higher angles of incidence, but nevertheless
the stopping fraction is, on average, higher than the sputtering yield for
target materials of all three types. The conclusion is that the `sputtering'
process usually results in an increase of the number of particles in a grain,
though not necessarily the grain mass.

It is vital to know whether embedded projectiles are likely to remain embedded
within the grain material, or whether they can escape on a timescale which
is very much shorter than the typical duration of the event which introduced
the sputtering. If an embedded projectile can escape quickly, then the
traditional sputtering yield is effectively equivalent to the true sputtering
yield, whilst if the projectile becomes permanently embedded, the true yield
is the difference between the traditional yield and the embedding fraction; this
difference may be negative, leading to grain growth.

Atoms in a solid can diffuse through the lattice, and the same mathematical
formalism can be used to describe this process as is used for gaseous
diffusion \cite{shewmon89}. The diffusion process may be used for both the
mobility of impurity atoms, and for `self-diffusion' of the elements present
in the lattice. Diffusion through solids is more complex than the analogous
process in gases because of the `quantized' nature of the motion, in which
an impurity atom must move from one lattice site to another. This extra
complexity is expressed through a diffusion `constant' which depends on the
packing of the lattice, its natural frequency of vibration, the temperature
of the lattice, and the binding energy of the impurity atom to the lattice.
The most important of these dependencies are the binding energy and temperature,
because they appear in an exponential Boltzmann factor: it is very difficult
for a strongly bound atom to diffuse through a very cold lattice, as expected.
The precise form of the equation for the diffusion coefficient depends upon
the mobility mechanism: for the vacancy mechanism, where an atom migrates by
jumping between randomly-placed lattice vacancies, the equation is
\begin{equation}
D(T) = a_{0}^{2} \nu \exp \left[\frac{S_{\nu}+S_{m}}{R}\right] \exp \left\{
   \frac{-(H_{\nu}+H_{m})}{RT}
                                                     \right\}
\end{equation}
where $a_{0}$ is the lattice spacing, $\nu$ is a directed oscillation frequency
(towards a vacancy), $R$ is the gas constant, $T$ is the grain
temperature, $S_{x}$ is an entropy and $H_{x}$ is an enthalpy. The subscript-$\nu$
contributions are from vacancies, and the $m$-subscript contributions refer
to lattice activation energies.
For the interstitial mechanism, where a small
atom `squeezes' through the lattice by deforming bonds, the equation is very
similar, but excludes the subscript-$\nu$ terms. A diffusion timescale is
$\tau_{diff} = d^{2}/(4\pi^{2} D(T))$ for spheres of diameter $d$ 
\cite{shewmon89}. From data in Norwick \& Burton \shortcite{nor75}, we can
deduce that at 100\,K, the measured temperature of the processed dust in
the SNR of Cas~A \cite{dunne03}, all the common elements will be fixed to their lattice
sites ($tau_{diff} > 10^{9}$\,yr) except H and He. For example, for carbon
in iron (assuming the plot in Norwick \& Burton can be extended down to
$100$\,K at the same gradient) the diffusion coefficient is of order $10^{-44}$\,cm$^{2}$\,s$^{-1}$, giving
a diffusion time many orders of magnitude longer than the age of the Universe,
for even the smallest grains.
The fate of the helium is
not crucial as it has a much lower abundance than H. For hydrogen, much depends on whether the H-atoms become chemically bound to the lattice. If they do
not, for example hydrogen in $\alpha$-iron \cite{nor75}, the diffusion
coefficient at $100$\,K lies between $10^{-9}$ and
 $10^{-6}$\,cm$^{2}$\,s$^{-1}$. Even assuming the slower value, hydrogen can
escape from a $10$\,nm layer on a timescale of $25$\,$\mu$\,s. Contrast
this with the diffusion of H$_{2}$ through a silicate, where chemical bonding
can take place. Here the diffusion coefficient at $100$\,K is $9.8\times10^{-27}
\,$cm$^{2}$\,s$^{-1}$, yielding a timescale of $80000$\,yr to escape a 
$100$\,nm layer. Hydrogen atoms are more reactive than molecules, and so would
be expected to remain longer in the solid. Chemically bound hydrogen can
survive for significant times in the solid.
These conclusions are of some importance in Section~3.4 and in Section~4.

Overall, typical deposition rates exceed sputtering rates. If the bombarding
gas has Galactic abundance, this will still lead to net erosion, since
sputtering by hydrogen and helium, even if these elements remain embedded,
results in irreplaceable loss of the metallic elements that form the grain
core. However, in metal-enriched gas, as found in supernova remnants, it is
possible that embedding of the metals can exceed the total rate at which
they are sputtered. Hydrogen and helium, if not chemically bound
to the lattice, can escape from a typical grain on timescales short 
compared to the time needed to erode a grain by sputtering. 

\begin{figure}
\vspace{0.6cm}
\psfig{file=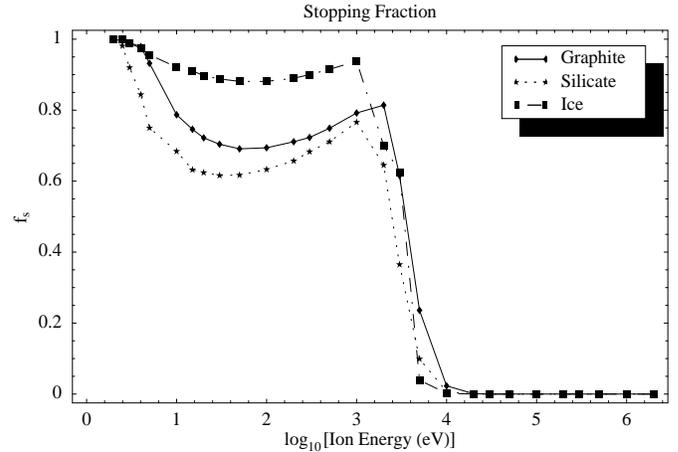,height=10.0cm,angle=270}
\caption{
The fraction of projectile ions stopped inside the target grain as a
function of projectile energy for zero angle of incidence. 
}
\label{f:stopper}
\end{figure}

\subsection{Processing of an olivine target}

Here, we present results from {\sc srim} computations which explicity take
into account the modification of a target by the combined processes of
sputtering and projectile stopping. The target material was olivine, as
in Section~3.1 and Section~3.2, but modification of the target grain was
studied in the following manner: Since {\sc srim} projectiles do not
modify the target with which they interact, the calculations were broken
down into a series of steps, each with a fixed target composition. Beginning
with a pristine target of olivine, {\sc srim} calculations were carried out
for all ten common projectile species at various angles and energies. Sputtering
yields and stopping fractions were recorded, and also the average penetration
depth for each projectile species, weighted by the distribution function
developed in Section~3.1. To begin the next step, the target was divided
into layers, according to the penetration depths, and the composition of
each layer modified to match a bombardment of 5\% by number of the original
target material. The modification by species was carried out in proportion
to the sputtering yields and stopping fractions for that species, and the
overall weight of each species determined according to Table~A1. This use
of direct abundances, rather than collision rates weighted by thermal
speeds of the individual species, compensates crudely for that fact that
the grains in the shocked gas would have a large initial drift velocity
with respect to the gas, which is only slowly dissipated.
A fresh
set of calculations was then performed for the modified target, which now
contains projectile species. New sputtering yields and stopping fractions
were computed, and the target modified again to match a further 5\% by
number of the original target. The penetration depths, however, were not
modified on the second or subsequent steps, since stopping distances have
very large standard deviations, and the number of differentiated target
layers would rapidly have reached unacceptable proportions.

\subsubsection{Calculations for the pristine target}

First, we report the penetration depths achieved for each projectile type
as a function of energy and impact angle. Since these values were also used
for all subsequent steps, we discuss them only once. The depth-data appear
in Table~4. The depths were originally measured at energies of
$10$, $30$, $100$, $300$, $1000$, $3000$ and $10000$\,eV, and angles of
$0$, $60$, $85$\degr, with the assumption of zero penetration at $90$\degr. In the columns of Table~4, the depths
have been averaged over solid angle, assuming an isotropic distribution of
impacts. The final column gives the grand-average over all the energies
used weighted according to the
 distribution function from eq.(\ref{eq:maxboltz}) and
eq.(\ref{eq:espec}).

\begin{table}
\caption{Angle-averaged penetration depths for projectile species incident on an
olivine target, as a function of energy. At $10$\,eV, all penetration
depths were negligible. Columns $2$-$6$ contain penetration depths at, respectively,
$30$, $100$, $300$, $1000$, $3000$ and $10000$\,eV. Column $7$ contains the
energy-averaged penetration depths.}
\begin{tabular}{@{}lrrrrrrr@{}}
\hline
Element&$d_{30}$&$d_{100}$&$d_{300}$&$d_{1k}$&$d_{3k}$&
        $d_{10k}$&$\bar{d}$\\
       &   nm      &     nm     &   nm       &   nm      &    nm     &
           nm       &  nm    \\
\hline
   H   & 0.7  &  1.5  &  3.2  &  8.4 &  20.6  & 45.6  & 1.746  \\
   He  & 0.4  &  0.9  &  1.9  &  5.4 &  14.0  & 35.2  & 1.036  \\
   C   & 0.3  &  0.4  &  0.8  &  1.7 &   4.0  & 11.4  & 0.472  \\
   N   & 0.3  &  0.4  &  0.8  &  1.6 &   3.5  &  9.8  & 0.472  \\
   O   & 0.3  &  0.4  &  0.7  &  1.5 &   3.2  &  8.7  & 0.444  \\
   Ne  & 0.3  &  0.4  &  0.7  &  1.4 &   2.8  &  7.5  & 0.444  \\
   Mg  & 0.3  &  0.4  &  0.7  &  1.2 &   2.6  &  6.7  & 0.446  \\
   Si  & 0.3  &  0.4  &  0.7  &  1.2 &   2.4  &  5.8  & 0.446  \\
   S   & 0.3  &  0.4  &  0.7  &  1.2 &   2.2  &  5.2  & 0.446  \\
   Fe  & 0.4  &  0.5  &  0.7  &  1.1 &   2.0  &  4.2  & 0.514  \\
\hline
\end{tabular}
\vspace*{1mm}
\noindent \\
\end{table}

When integrating over the distribution function, we have
adopted the following standard parameters: shock speed, $u_{100}=4.0$,
Alfv\'{e}n speed, $v_{A10}=1.0$, and ratio of specific heats in the
thermal gas, $\gamma_{g}=5/3$. These parameters are typical of a supernova
blast wave, close to its transition from the Sedov-Taylor phase to the
pressure-driven snowplough phase of the remnant.
In Table~5, we display the sputtering
results for the pristine target. Each row represents a 
different type of projectile; the first four columns are sputtering yields
for the four elemental consituents of the olivine; the
final column is the stopping fraction. The sputtering yields and the stopping
fraction
are averaged over angle and over impact energy, weighted by the
distribution function. However, the yields are not corrected for either
the stoichiometric ratio of oxygen in the target, or for the relative
abundances of the projectile elements.
Table~5, together with Table~4 and Table~A1,
provide the means to alter the target composition in line with the
expected bombardment in the post-shock regime.

\begin{table}
\caption{Angle- and energy-averaged sputtering yields for the ten common
projectile species incident on a pristine olivine target. The final
column is the angle- and energy-averaged stopping fraction for that
projectile.}
\begin{tabular}{@{}lrrrrr@{}}
\hline
Element&$Y_{Mg}$&$Y_{Fe}$&$Y_{Si}$&$Y_{O}$&$f_{stop}$ \\
       &  \%    &  \%    &  \%    &  \%   &           \\
\hline
   H   & 1.06    & 0.73    & 0.71    & 3.78    & 0.435  \\
   He  & 5.86    & 2.96    & 2.90    & 19.7    & 0.494  \\
   C   & 13.2    & 7.35    & 7.03    & 45.5    & 0.642  \\
   N   & 13.8    & 7.68    & 7.28    & 47.3    & 0.661  \\
   O   & 14.1    & 7.89    & 7.44    & 48.5    & 0.677  \\
   Ne  & 14.9    & 8.21    & 7.71    & 50.8    & 0.698  \\
   Mg  & 15.2    & 8.37    & 7.79    & 51.7    & 0.714  \\
   Si  & 14.9    & 8.10    & 7.43    & 50.5    & 0.735  \\
   S   & 14.4    & 7.87    & 7.06    & 49.0    & 0.753  \\
   Fe  & 13.6    & 6.64    & 6.35    & 45.1    & 0.803  \\
\hline
\end{tabular}
\vspace*{1mm}
\noindent \\
\end{table} 

\subsubsection{Calculations for the modified target}

Following the calculation on the pristine target, the target material was
divided into layers, dependent on penetration depth. Each projectile
material was assumed to be uniformly distributed between the surface and
its own penetration depth, $\bar{d}$ from Table~4, and to have zero abundance at any greater depth
in the target. Although, in principle, the modified target would have
eleven layers (unmodified olivine plus ten layers modified by successive
mixtures of projectile species) the average penetration depth data (final
column of Table~4) allows a simplification in which the target is represented
by unmodified olivine plus five modified layers. The outermost layer is
contaminated by all the projectile species, the second by H, He, C, N and Fe,
the third by H, He and Fe, the fourth by H and He, and the fifth by
hydrogen alone. Each projectile species was assumed to sputter target atoms
uniformly from all the zones into which that species could penetrate.

\begin{table}
\caption{Final abundances in the grain, after 35\% bombardment, for all penetrated layers, as 
a fraction of the total number of nuclei present in the layer, for 
the ten most common elements.}
\begin{tabular}{@{}lrrrrr@{}}
\hline
Element& Layer $1$ & Layer 2 & Layer 3 &Layer 4 & Layer 5 \\
\hline
   H   & $0.1228$   & $0.1227$  & $0.1228$ & $0.1228$ & $0.1250$ \\
   He  & $0.0219$   & $0.0219$  & $0.0219$ & $0.0219$ & -        \\
   C   & $2.41(-4)$ & $2.24(-4)$ & -        & -        & -        \\
   N   & $8.0(-5)$  & $6.8(-5)$  & -        & -        & -        \\
   O   & $0.4870$   & $0.4869$  & $0.4871$ & $0.4871$ & $0.4993$ \\
   Ne  & $1.17(-4)$ & -         & -        & -        & -        \\
   Mg  & $0.1216$   & $0.1216$  & $0.1216$ & $0.1216$ & $0.1248$ \\
   Si  & $0.1231$   & $0.1232$  & $0.1233$ & $0.1234$ & $0.1254$ \\
   S   & $1.7(-5)$  & -         & -        & -        & -        \\
   Fe  & $0.1232$   & $0.1233$  & $0.1234$ & $0.1233$ & $0.1260$ \\
\hline
\end{tabular}
\vspace*{1mm}
\noindent \\
\end{table}

Bombardment of the target was carried out in a sequence of steps, with the
first step being bombardment of the pristine target. Throughout each step
the composition of the target, in a given layer, was assumed to be constant.
At the end of each step, the composition of the target was adjusted to take
into account both the loss of sputtered atoms, and the embedding of stopped
projectiles. At all steps later than the first, the projectile species form
embedded populations, which may themselves be sputtered from the target.
The very low abundance of all projectiles heavier than He made it possible
to ignore the sputtering effect of carbon and heavier elements, even though
these heavy nuclei are individually more efficient sputterers then 
hydrogen and helium (see Table~5). Although ignored as sputtering projectiles, the
embedding of the heavier elements was still considered, and populations of
these species were also subject to sputtering by H and He. For the effects
of heavy elements in conditions with non-standard abundances (in SNRs) see
Section~4.

\begin{figure}
\vspace{0.6cm}
\psfig{file=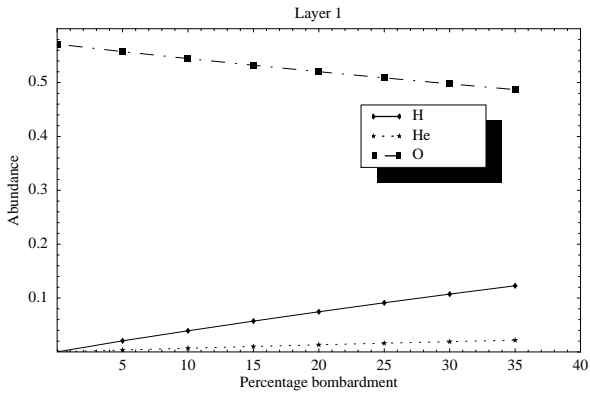,height=6.0cm,bbllx=100pt,bblly=-400pt,bburx=282pt,bbury=-50pt,angle=0}
\caption{
Changes in the abundances of hydrogen and helium projectiles and oxygen in
a 100\,nm grain as a function of total bombardment for the outermost layer
(see text for definitions). 
}
\label{f:layer1a}
\end{figure}

\begin{figure}
\vspace{0.6cm}
\psfig{file=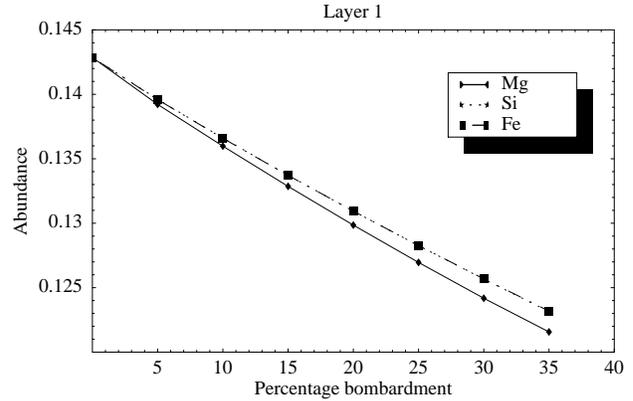,height=6.0cm,bbllx=100pt,bblly=-350pt,bburx=282pt,bbury=-70pt,angle=0}
\caption{
Changes in the abundances of iron, silicon and magnesium in
a 100\,nm grain as a function of total bombardment for the outermost layer
(see text for definitions). The graphs for iron and silicon are so close
together that they are virtually indistinguishable.
}
\label{f:layer1b}
\end{figure}

\begin{figure}
\vspace{0.6cm}
\psfig{file=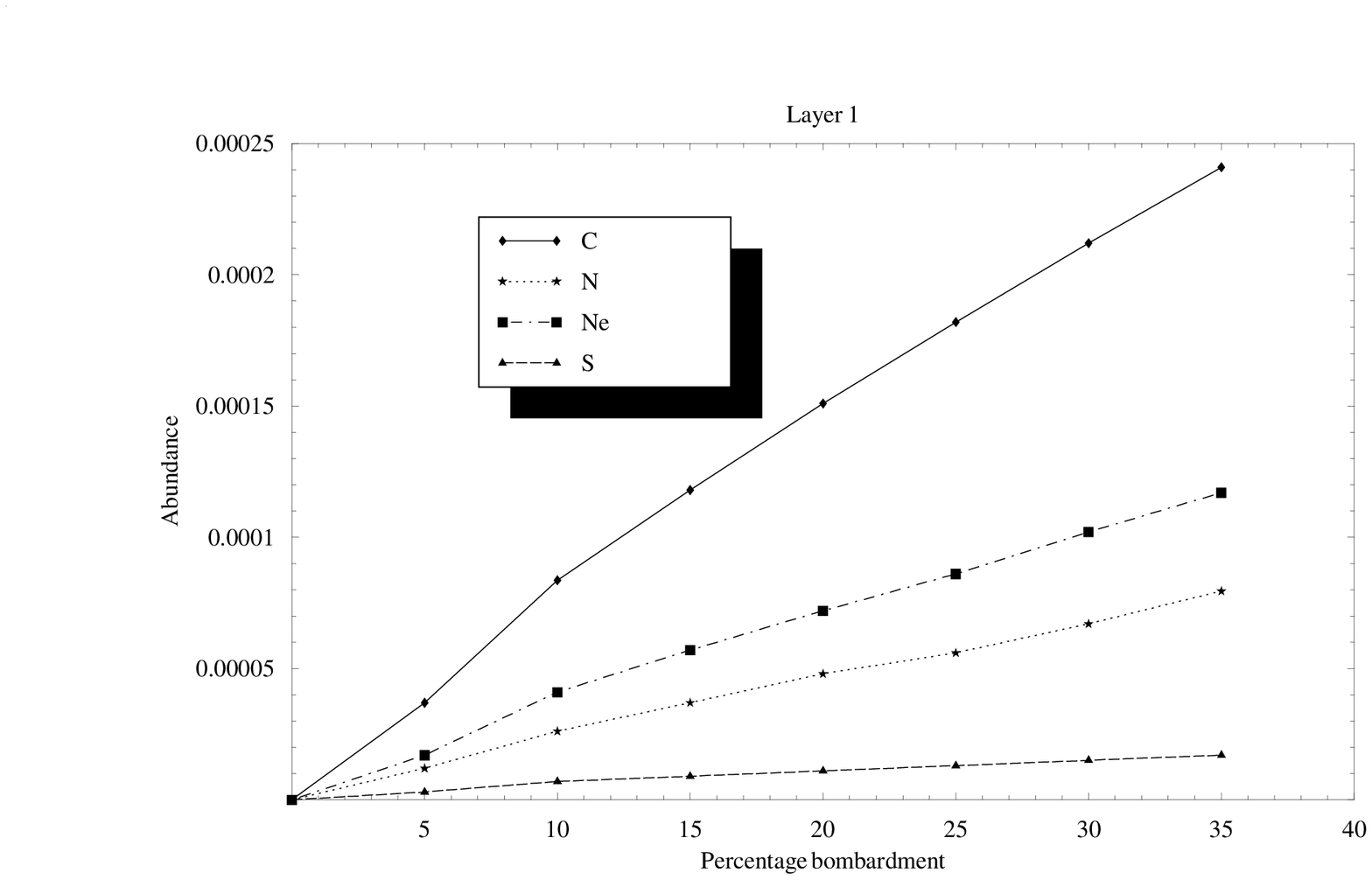,height=3.5cm,bbllx=180pt,bblly=110pt,bburx=380pt,bbury=310pt,angle=270}
\caption{
Changes in the abundances of the `rare', heavy projectiles, C,N,Ne and
S in
a 100\,nm grain as a function of total bombardment for the outermost layer
(see text for definitions).
}
\label{f:layer1c}
\end{figure}

The sequence of graphs, Fig. 7 to Fig. 9, shows
the abundances of all species in the outermost layer of the grain as a function
of bombardment. The level of
bombardment is expressed as a percentage, which is $100\times$ the total
number of projectile atoms divided by the total number of vulnerable target
atoms (summed over all five layers) in the pristine target.
This degree of bombardment can, of course, be converted to
time, given a particular astrophysical environment; we consider the effect
of supernova shocks below. Fig. 7 shows the rising abundances of the sputtering
projectiles, hydrogen and helium, together with the decrease in the fraction
of oxygen, the most abundant element in the pristine target. Oxygen is the
most easily displaced of the elements initially present in the olivine. In
Fig. 8, we show the falling abundances of the metallic components of the
olivine: Fe, Si and Mg. Magnesium is significantly more easily sputtered than
the other two elements, and we note that the sputtering of Si and Fe is so
similar that the graphs for these two elements are indistinguishable in
the figure. Fig. 9 shows the rising abundances of those projectiles which
are not considered to cause significant sputtering. The low cosmic
abundance of these elements means that small contaminant populations develop
in the outer regions of the grain.

Overall, the effect of the sputtering process is not so much to erode the
grain as to process it by significantly changing its composition. In all the
penetrated zones, we find that the total zone mass and the zone density fall
below the values set for the pristine olivine target, but that the particle
number in all zones actually rises if we assume that H and He projectiles
remain embedded in the target. We have shown in Section~3.3 that long-term
survival of at least some hydrogen is likely in a silicate, providing it
can bind to the lattice. The large number of vacancies produced by the
sputtering process, leaving a general defficiency of oxygen, will aid
binding by providing many lattice sites with low energy barriers. The low
abundance of He, compared to H, makes its loss or retention in the grain of
little importance.  

\section{Survival of grains against supernova shockwaves}

Recent observational studies with SCUBA on the JCMT \cite{dunne03}
have convincingly
demonstrated that supernovae (at least those of type II) are net producers
of dust. We therefore proceed in the spirit of attempting to explain this
observational fact, rather than attempting to make
our models agree with previous work, which tends to favour net erosion by
sputtering.

The first point that should be made is that a dust grain will experience many
collisions with gas atoms and ions in a typical remnant. In most cases, the
grain will receive easily enough bombardment to completely erode it if the
collisions lead to net sputtering over embedding. The mean-free time in
seconds for collisions between gas atoms/ions and
a dust grain in the post-shock gas, is given by
\begin{equation}
\tau_{coll} = 4.087\times 10^{5}
              \frac{T_{1000}^{3/8} \mu^{1/2}}
                   {n_{cc} v_{100}^{3/4} T_{5}^{1/2} a_{nm}^{2}} \; \; s
\end{equation}
where $T_{1000}$ is the pre-shock gas temperature in units of $1000\,K$, $\mu$
is the relative mass of the bombarding particle relative to H, $n_{cc}$ is
the number density of this bombarding species in units of cm$^{-3}$, v$_{100}$
is the shock speed in units of $100$\,km\,s$^{-1}$, $T_{5}$ is the post-shock
temperature in units of $10^{5}$\,K, and $a_{nm}$ is the grain radius in
nm. In deriving this equation, we have used the definition of the compression
ratio from Berezhko \& Ellison \shortcite{berez99}, which is
$n_{2}/n_{0} = 1.3 (v_{s}/c_{s2})^{3/4}$, where $v_{s}$ is the shock speed
and $c_{s2}$ is the sound speed in the pre-shock gas. For our model grain
in Section~3.4, being bombarded by hydrogen, the timescale for $100$\%
bombardment (being struck by a number of H-nuclei equal to the total number of
atoms in the vulnerable part of the grain) is only 
\begin{equation}
\tau_{100} = 24.4 \frac{T_{1000}^{3/8}}
                       {n_{cc} v_{100}^{3/4} T_{5}^{1/2}} \;\;yr
\end{equation}
This figure is largely size independent down to the size where the vulnerable
zone becomes equal to the entire grain, since both the volume of the vulnerable
zone and the grain cross-section scale as $a^{2}$. To sputter out all the
metallic elements from the vulnerable zone requires a few times $\tau_{100}$,
or about a century. For larger grains, including our example, we can view
grain destruction as requiring the removal of several sequential layers, but
even for our $100$\,nm grain, total destruction would occur on a timescale
less than $1000$\,yr; optimistic choices of $n_{cc}$ yield $\sim 10^{4}$\,yr.
The fact that total particle number initially rises
cannot prevent this destruction: the helium, which anyway has a low abundance
within the grain, would probably diffuse out of the sample in a shorter time
than $\tau_{100}$; hydrogen certainly would if it did not bind chemically to
the lattice. Although much H might bond, hydrogen which did not attach
chemically could escape on a very short timescale (see Section~3.3). Therefore
the true abundance of H must always be less than in our standard model, and
as metals are lost, the chances of forming a chemical bond, and becoming
part of the grain material, are progressively reduced.

In a supernova remnant, abundances are very different to those in our standard
model if we are considering material which is part of the ejecta, or material
which is part of the ISM in a young remnant: the early swept material is almost
certainly stellar wind debris, which contains enhancements of the elements He,
C, N, and O with respect to the Galactic norm. For the ejecta, we have
used abundances computed by Limongi \& Chieffi \shortcite{lim03}. Using the
Salpeter IMF, the mean mass of a supernova progenitor is $18.7$\,M$_{\odot}$
if the upper mass limit is set to $100$\,M$_{\odot}$, but $15.0$\,M$_{\odot}$
if we limit the upper progenitor mass to $40$\,M$_{\odot}$. We have
selected the latter figure, and abundances, relative to hydrogen, appear in
the right-hand column of Table~A1. To obtain our tabulated values, we have
summed over all isotopes used by Limongi \& Chieffi, and averaged over the
six models computed for the mass of $15$\,M$_{\odot}$. To proceed further,
we consider first the fate of the supernova ejecta, which are shocked by
the reverse shock, and then the wind debris and raw ISM, which are shocked
by the external blast-wave.

\subsection{Dust in the supernova ejecta}

Initially, the ejecta are cold and unshocked, with a contact discontinuity
separating them from material swept-up by the blast-wave \cite{true99}. Even
when a reverse shock is set up by the overpressure of the hot swept material,
the cooling times in the ejecta are very short, leading to an early radiative
phase in the ejecta \cite{true99}. Dust can condense under these conditions
within a very short time, compared to the evolutionary scale-time of the
free-expansion phase of the remnant (of order 1000\,yr or less). There is
observational evidence for dust condensing in SN1987A within 530\,d of the
outburst \cite{dwek98}. As the reverse shock accelerates, higher temperatures
are attained in the shocked ejecta, but though the bombardment is efficient,
the ejecta are metal-rich. Although data in Table~5 show sputtering
exceeding embedding, for metallic elements, by a factor of $\sim 3$, it is
quite possible that net embedding could predominate. Values of $E_{B}$, $E_{D}$
and $E_{s}$ closer to the May et al. \shortcite{may00} averages would lower
sputtering without significantly reducing embedding rates. Also, the 
deposition of large amounts of carbon and hydrogen in a thin
surface layer are likely to lead to
the formation of a tough organic mantle on grains initially of silicate
composition \cite{gal96} \cite{li97}. To compute the sputtering yields from such a mantle, as a function of impact energy and time, requires much additional work
but such a mantle is likely to be at least as resistant as carbon, with a yield of
about a factor of three lower than for silicates (see Figure~2). The fact that supernovae
appear to be net producers of dust from observations means that the
investigation of such mantle formation is worthwhile. Protection of the
underlying silicate could be effective
until typical collision energies
reach 300-400\,eV, allowing the stopping of metallic projectiles to exceed their losses in
sputtering. Net growth of the freshly condensed grains could therefore
continue
in the ejecta until roughly the end of the free-expansion phase. Truelove \&
McKee \shortcite{true99} quote a characteristic shock speed for the reverse
shock of,
\begin{equation}
v_{ch} = 7090 E_{51}^{1/2} (M_{ej}/M_{\odot})^{-1/2} \; \; km\,s^{-1}
\end{equation}
for supernova energy $E_{51}$, in units of $10^{51}$\,ergs, with $M_{ej}$
solar masses of ejecta. The end of the free-expansion phase occurs when
the speed is $0.585$ times the characteristic value. The mean values of
$E_{51}$ and $M_{ej}$ for the $15$\,M$_{\odot}$ model in Limongi \& Chieffi \shortcite{lim03} are , respectively, $0.877$ and $13.43$, giving a shock
speed of $1060$\,km\,s$^{-1}$. This corresponds, via eq.(\ref{eq:pstemp}),
to a temperature of $3.73\times 10^{6}$\,K, and an energy of $320$\,eV.
The mass of shocked ejecta, at this same stage of evolution, is $0.781M_{ej}$,
so most of the ejecta will form dust, may accumulate metals, and never be
shocked to a high enough energy for net sputtering to dominate. Even the
remaining $22\%$ of the ejecta, which is likely to lose its newly-condensed
dust to sputtering, will form metal-rich gas which, when it eventually
cools, is likely to add to the dust already present in the cooler shocked
ejecta. We note that the cooling time is given by
\begin{equation} \label{eq:tcool}
\tau_{cool} = 6.3 \times 10^{-5} \frac{T^{3/2}}{\zeta_{m} n_{cc}} \; \; yr
\end{equation}
where $n_{cc}$ is the number density of the cooling gas in cm$^{-3}$ and
$\zeta_{m}$ is a metallicity correction factor \cite{cio88}. Therefore, the
destruction of the dust enhances $\zeta_{m}$, and enhances cooling towards
temperatures where the bombardment process results in net grain growth.
Overall then, much of the metal content of the ejecta can survive the
supernova remnant bound in dust. We note that the survival of initially
condensed ejecta dust was predicted by Barlow \& Silk \cite{bas77}, but this
would provide insufficient mass to make supernovae net creators of dust.

\subsection{Dust in the ISM and wind debris}

The fate of grains in the wind debris and raw interstellar gas is more
complicated. We expect the wind material to have enhanced abundances (over
the mean ISM values) of lower-mass nucleosynthesis products, but not to
have high abundances of the iron-group elements. We also expect successive
shells of wind debris to have come from a sequence of evolutionary stages
with different chemical composition. Nevertheless, we expect much of this
material to resemble red supergiant winds, and to be rich in dust. For
most reasonable parameter values, the entire wind-zone will be swept within the free-expansion
phase of the remnant, and will therefore be shocked and heated to very high temperatures. For the particular case of the mean Limongi \& Chiefi
\shortcite{lim03} $15$\,M$_{\odot}$ model, the blast-wave speed at the
end of the free-expansion phase is still $1665$\,km\,s$^{-1}$ which is
destructive, even for metal-rich gas. However, as in the inner regions of
the ejecta (see above) the dust destruction leads to metal-rich gas with
cooling times that are much more rapid than those for typical ISM gas.
With the help of the Berezhko \& Ellison \shortcite{berez99} formula,
$n_{2}/n_{0} = 1.3 (v_{s}/c_{s2})^{3/4}$, we can calculate a compression factor of $113$ for
the end of the free-expansion phase, and a corresponding temperature of
$6.1\times 10^{6}$\,K from eq,(\ref{eq:pstemp}). With typical coolants
at levels of $100$ times their typical ISM abundances, eq.(\ref{eq:tcool}) 
reduces to $84/n_{cc}$\,yr which is of similar order to the grain
destruction time. It is therefore likely that the shell of wind debris
becomes rapidly radiative, and that dust either survives or rapidly
recondenses, joining the mass of dust formed in the ejecta. This mass of
dust is likely to be similar to, but not greater than, the mass of dust
in the ejecta, on the grounds that the pre-supernova star was unlikely to
have lost more than half its ZAMS mass as wind debris.

Once the blast-wave has cleared the wind material, usually before the end
of the free-expansion phase of the remnant, it begins to sweep-up
unenriched ISM gas, which is likely to contain dust at typical ISM
proportions of $\sim$1\% by mass. This gas is still heated to high
temperatures, and now there is no chance of net embedding for elements
heavier than H and He (see Section~3.4). Gas bombardment of grains in
this regime can only be destructive, and nor will the release of metals
into the gas reduce the cooling time very much. This state of affairs will
last until the end of the Sedov-Taylor phase, that is until the shock
becomes radiative, and cooling times, even for unenriched gas, become
short compared to the evolutionary timescale of the remnant. Some
dust destruction may continue into the early PDS (pressure-driven
snowplough) phase which follows the Sedov-Taylor era. 

A rough estimate of the amount of dust destroyed can be obtained from the
mass swept by the blast-wave during the Sedov-Taylor (ST) phase of the
SNR. The PDS phase
 begins after $\sim 13000 n_{0cc}^{-4/7}\zeta_{m}^{-5/14}$\,yr
\cite{cio88}. From the end of the free-expansion phase up to this time,
the remnant sweeps up a mass of
\begin{equation} \label{eq:stmass}
M = 17.5 \bar{\mu} (M_{ej}/M_{\odot}) \; \; M_{\odot}
\end{equation}
which is notably independent of the ISM density. The quantity
$\bar{\mu}$ is the mean relative particle mass in the unshocked ISM.
For our standard model at
$15$\,M$_{\odot}$, we find a total swept mass of $306$\,M$_{\odot}$, so for
a typical dust mass fraction about $3$\,M$_{\odot}$ of 
pre-supernova dust is destroyed.

\section{Discussion}

The results of this work are broadly as follows: Firstly, sputtering in the
classical sense is almost invariably an inefficient process: In most
situations, the sputtering yield for a hydrogen projectile, and the chosen
targets, is far lower than the 100\% required to ensure erosion
of the grain. Slow erosion is still possible at sputtering yields below
100\%, but this requires that most of the incident projectiles do not
get stopped in the target and hence become part of the grain. In fact, we
find that the stopping fraction is typically large, and
ususally,
the number of atoms in the grain will increase due to particle collisions, as
a result of the stopping of projectile atoms within the grain. In many cases,
the grain mass will actually increase. For particle collisions to be effective
at destroying grains, we need to appeal to the secondary effects of lightening
the average atomic mass in the grain, the disruption of its crystal
structure and diffusion of light elements from the grain.

The analysis of survival of dust in a supernova remnant produces an ambiguous
result, but we cannot exclude net production on the basis of this very
simple estimate. The Sedov-Taylor phase destroys $\sim 3$\,M$_{\odot}$ of
ISM dust. On the other side of the balance sheet, the
mass of elements heavier than He in the ejecta is $\sim1.2$\,M$_{\odot}$, so
the ejecta and wind debris could between them contribute about 
$2$\,M$_{\odot}$ of dust. The masses of dust produced and destroyed are
very similar, and it is necessary to complete a very detailed analysis
to find out which predominates, though it appears to be a requirement from
models of galactic dust evolution \cite{mike01} that supernovae must be net producers of dust.

The dust produced by condensation followed by bombardment would have an
interesting structure: the core would be a condensate, possibly crystalline
initially, but bombardment would then insert heavy elements at random down
to their typical penetration depths in the grain. Thus even if the original
condensate were a silicate, the outer layers would be rich in heavy elements
that did not form part of the mineral: carbon and nitrogen for example.
This formation and processing scheme could make grains very similar to
the silicate core with `organic' mantle type that form the large grain
population in Greenberg's unified model \cite{gal96} \cite{li97}. The dust
is also formed in the correct region, as observationally determined:
between the blast-wave and reverse shocks \cite{dwek98}. 
Changes in dust grain structure due to bombardment could also affect the
optical properties of the grains, and this may be required to explain the
SCUBA observations of supernova dust emission compared with more general
ISM dust.

\subsection*{ACKNOWLEDGMENTS}

MDG acknowledges the award of
a University Research Fellowship by the Royal Society.

\appendix

\section{Abundances}

In our standard model,
projectile ions were assumed to have Galactic abundances, and the ten
most common elements were used, with abundances, 
relative to hydrogen, listed in the second column of Table~A1. All other elements, at Galactic abundance, have
individual abundances lower than $3.58 \times 10^{-6}$ relative to
hydrogen.

\begin{table}
\caption{Galactic Abundances}
\begin{tabular}{@{}lrr@{}}
\hline
Element   &  Galactic              &   SN ejecta           \\
\hline
   H      &  $1.0$                 &  $1.0$                \\
   He     &  $0.0975$              &  $0.703$              \\
   C      &  $3.62 \times 10^{-4}$ &  $2.89\times 10^{-2}$ \\ 
   N      &  $1.13 \times 10^{-4}$ &  $7.10\times 10^{-3}$ \\
   O      &  $4.91 \times 10^{-4}$ &  $7.30\times 10^{-2}$ \\
   Ne     &  $1.23 \times 10^{-4}$ &  $8.79\times 10^{-3}$ \\
   Mg     &  $3.84 \times 10^{-5}$ &  $5.87\times 10^{-3}$ \\
   Si     &  $3.58 \times 10^{-5}$ &  $1.66\times 10^{-2}$ \\
   S      &  $1.85 \times 10^{-5}$ &  $7.70\times 10^{-3}$ \\
   Fe     &  $3.15 \times 10^{-5}$ &  $2.63\times 10^{-2}$ \\
\hline
\end{tabular}
\end{table}

The right-hand column of Table~A1 shows abundances in supernova ejecta
computed by Limongi \& Chieffi \shortcite{lim03} for a $15$\,M$_{\odot}$
progenitor. These figures are the sums over all isotopes shown, and
averaged over the six models run for this mass.

\end{document}